\documentclass[12pt,a4paper]{scrartcl}
\usepackage[utf8]{inputenc}
\usepackage{booktabs}
\usepackage[table,dvipsnames]{xcolor}
\usepackage{textcomp}
\usepackage{stackengine}
\usepackage{arydshln}
\usepackage{graphicx}
\usepackage{setspace}
\usepackage{tabularx}
\usepackage{tcolorbox}
\usepackage{booktabs}
\usepackage{threeparttable}
\usepackage{supertabular}
\usepackage{rotating}
\usepackage[longnamesfirst]{natbib}
\usepackage[]{csquotes}
\usepackage{amsmath}
\usepackage{url}
\usepackage{tikz}
\usetikzlibrary{tikzmark,matrix,calc}
\usepackage{amssymb}
\usepackage{eurosym}
\usepackage{multirow}
\usepackage{environ}
\usepackage[margin=2.5cm]{geometry}
\usepackage[multiple]{footmisc}

\title{Very Highly Skilled Individuals Do Not Choke Under Pressure: Evidence from Professional Darts} 

\author{
Christian Deutscher\footnote{Corresponding author; email: \texttt{christian.deutscher@uni-bielefeld.de}.},
Marius \"Otting, 
Roland Langrock, \\
Sebastian Gehrmann,
Sandra Schneemann, Hendrik Scholten\\[1em] 
Bielefeld University, Germany}

\date{}

\begin{document}
\maketitle

\begin{abstract} 

Understanding and predicting how individuals perform in high-pressure situations is of importance in designing and managing workplaces, but also in other areas of society such as disaster management or professional sports. For simple effort tasks, an increase in the pressure experienced by an individual, e.g.\ due to incentive schemes in a workplace, will increase the effort put into the task and hence in most cases also the performance. However, for skill tasks, there exists a substantial body of literature that fairly consistently reports a \textit{choking} phenomenon, where individuals exposed to pressure do in fact perform worse than in non-pressure situations. However, we argue that many of the corresponding studies have crucial limitations, such as neglected interaction effects or insufficient numbers of observations to allow within-individual analysis. We also diagnose some degree of confusion in the literature regarding the differentiation between skill and effort tasks, which, assuming the choking phenomenon in case of skill tasks to be real, would in fact be assumed to be orthogonally impacted by pressure. Focusing on the more complex and usually harder to capture case of skill tasks, here we investigate performance under pressure in professional darts as a near-ideal setting with no direct interaction between players and a high number of observations per subject. We analyze almost one year of tournament data covering 23,192 dart throws, hence a data set that is very much larger than those used in most previous studies. We find strong evidence for an overall improved performance under pressure, for nearly all 83 players in the sample. Contrary to what would be expected given the evidence in favor of a choking phenomenon, we hence find that professional darts players \textit{excel} at performing skill tasks under high pressure. These results could have important consequences for our understanding of how highly skilled individuals deal with high-pressure situations. 

\vspace{0.3cm} \noindent \textbf{JEL-Code}: D91, M50

\vspace{0.2cm} \noindent \textbf{Keywords}: performance under pressure; behavioral economics; sports economics; skill task

\end{abstract}

\doublespacing

\section{Introduction}
The effect of pressure on human performance is relevant in various areas of the society, including sports competitions \citep{hill2010choking}, political crises \citep{boin2016politics}, and performance-based payment in workplaces \citep{ariely2009large}, to name but a few. A broad distinction differentiates between effort and skill tasks. Success in effort tasks is dependent on motivation to perform while skill task outcome underlies precision of (often automatic) execution. For effort tasks, such as counting digits \citep{konow2000fair} or filling envelopes \citep{abeler2011reference}, 
individuals will typically respond to increased pressure (e.g.\ resulting from performance-related payment schemes) by investing more effort, which given the nature of such tasks will improve their performance \citep{lazear2000, paarsch1999, paarsch2000, prendergast1999provision}. However, there is broad agreement in the literature that performance in skill tasks, e.g.\ juggling a soccer ball \citep{ali2011measuring}, declines in high-pressure or decisive situations, commonly referred to as ``choking under pressure''. According to \citet{beilock2007athletes}, an individual is choking under pressure when their performance is worse than expected given their capabilities and past performances. While there may also be random fluctuations in skill levels, choking under pressure refers to systematic suboptimal performance in high-pressure situations.

Choking under pressure can be caused by changes in the execution of actions, or simply distraction, generated either by rewards in case of success \citep{baumeister1984choking,ariely2009large} or potential penalties in case of failure \citep{kleine1988anxiety}. Empirical findings related to performance under pressure --- both such that are based on experimental data but also those using field data --- consistently confirm a negative impact of pressure on skill tasks. Studies using field data often focus on sports, e.g.\ penalty kicks in soccer, free throws in basketball, or putting in golf. While the existing literature seems to provide overwhelming evidence in favor of the choking under pressure phenomenon, we believe that most of the existing studies have crucial limitations, most notably neglected interaction effects between competitors and misleading benchmarks for performance in non-pressure situations. Second, reported results often mix effort and skill tasks despite the projected contrary impact of pressure on performance. 

Here we use a large data set from professional darts, comprising 23,192 individual dart throws, for a comprehensive empirical test of the choking under pressure theory. For the professional darts player analyzed in this study, playing darts is a full time job. The top players regularly earn prize money exceeding one million Euro per year. In professional darts, highly skilled players repeatedly throw at the dartboard from the exact same position effectively without any interaction between competitors, making the task highly standardized. The high level of standardization of individual throws as well as the very many repetitions of almost identical actions, performed by professionals, renders darts a near-ideal setting for measuring the effect of pressure on performance. The amount of data available on throwing performances not only allows for comprehensive inference on the existence and the magnitude of any potential effect of pressure on performance, but also enables to track the variability of the effect across players. As pointed out by \citet{mcewan2013effects}, throwing darts is a skill task as it refers to a highly standardized movement which requires a high motor skill in order to perform well. Hence, following the literature, performance in darts would be expected to decline as pressure increases. Against the current state of research on motor tasks, we find (nearly all) professional darts players to excel in pressure situations. We argue that highly skilled individuals are able to digest pressure situations towards a positive outcome, while those who choke under pressure decide towards other professions. This emphasizes the importance of possession of skills for the impact of pressure on performance.

The paper is structured as follows. Section \ref{chap:Performance and Choking Under Pressure} reviews the literature on performance under pressure, and in particular details what we consider to be crucial limitations of existing studies. In Section \ref{chap:Pressure Situations in Darts}, we explain the rules of darts and define what constitutes a pressure situation in darts. Section \ref{chap:Empirical Analysis} presents the empirical approach and results. Section \ref{chap:Throwing Darts - Skill or Effort Task?} provides further indication that throwing darts is a skill task.

\section{Performance and Choking Under Pressure}\label{chap:Performance and Choking Under Pressure} 

\subsection{Terminology}
Pressure results from individuals' ambitions to perform in an optimal way in situations where high-level performance is in demand \citep{baumeister1984choking}. Performance under pressure could in principle go either way, i.e.\ high expectations towards (the own) performance could impact performance in a negative or a positive way (or not at all). To measure the impact of pressure, performance in pressure situations is compared to performance in non-pressure situations. Choking under pressure refers specifically to a \textit{negative} impact of high performance expectations \citep{baumeister1986review,hill2009re}.

\subsection{Theories on Impact of Pressure}
The impact of pressure on performance crucially depends on the type of task to be performed. Tasks can be such that performance is determined mostly by effort, or alternatively a task can be such that the skill level is the key factor for success. For effort tasks, pressure situations result in increased effort and hence improved performance \citep{rosen1986prizes}. For skill tasks, performance has repeatedly been demonstrated to be impaired by pressure, which is commonly referred to as choking under pressure \citep{wallace2005audience}. While the effect of pressure on effort tasks is obvious, in skill tasks the potential psychological factors at play are likely more complex, such that we focus on these tasks in the following.

Choking under pressure in skill tasks may be related to various drivers. In particular, different skills may make use of different memory functions, namely explicit and procedural memory, respectively \citep{beilock2010}. Explicit memory enables the intentional recollection of factual information, while procedural memory works without conscious awareness and helps at performing tasks. Two classes of attentional theories capture choking under pressure, distraction theories and explicit monitoring theories \citep{decaro2011choking,hill2010choking}.\footnote{Some authors argue that distraction and explicit monitoring theories are not necessarily mutually exclusive, but rather complementary \citep[see e.g.][]{beilock2001on, sanders2012shirking}.}
Distraction theories claim high-pressure situations to harm performance by putting individuals’ attention to task irrelevant thoughts \citep{beilock2001on,lewis1997thinking}. Put in a nutshell, individuals concern about two tasks at once, since the situation-related thoughts add to the task to be performed. Given the restricted working memory individuals performance declines as focus is drawn away from the task \citep{engle2002working}.

On the other hand, self-focus or explicit monitoring theories explicitly predict that pressure increases self-consciousness to a point where it harms performance (overattention). It can cause the skilled performer to deviate from routine actions or, put differently, to utilize the procedural memory less \citep{markman2006choking}. Instead, closer attention is paid to the processes of performance and its step-by-step control. This ties in with the concept of skill acquisition: when initially learning a skill, performance is controlled consciously by explicit knowledge as skills are executed step-by-step \citep{anderson1982acquisition}. Over time and through practice, skills become internalized and usage of conscious control decreases. Pressure can interfere with this now automated control processes of skilled performers \citep{wulf2007external}. Under pressure, actions are no longer executed automatically as attention is redirected to task execution \citep{decaro2011choking}. The overall sequence of actions is broken down into step-by-step control as in early stages of learning, resulting in impaired performance \citep{masters1992knowledge}. Consequently, individuals consciously monitor and control a skill they would perform automatically in non-pressure situations \citep{jackson2006attentional,decaro2011choking}.

\subsection{Empirical Findings for Performance under Pressure in Skill Tasks}
As this paper analyses performance under pressure in a sport-related skill task, this section is devoted to previous findings from sports.\footnote {Early non-sport studies include \citet{baumeister1984choking, heaton1991self}.} 
In an experimental setting \citet{lewis1997thinking} discover golf putting performance to be worse when subjects are put under pressure. However, in high-pressure situations participants who are distracted by a secondary task (counting down from 100) outperform subjects who solely concentrate on the putting task. The latter result is explained by too much focus on the task execution induced by the additional motivation to perform well in high-pressure conditions. The additional focus disturbs task execution which normally is performed automatically. Further evidence for diminishing golf putting performance under pressure is presented by \citet{beilock2001fragility} who ask 108 undergraduate students with little or no golf experience to putt a golf ball as close to a target as possible. Considering different kinds of intervention methods, the authors create pressure-like situations using monetary incentives. Results generally confirm decreasing performance for high-pressure situations. However, the authors show putting accuracy to slightly increase under pressure when subjects had made their practice putts under self-consciousness-raising conditions.

For a hockey dribbling task with 34 experienced participants, \citet{jackson2006attentional} find that performance is worse in high-pressure situations. Results further show that within high and low-pressure conditions subjects perform better when not concentrating explicitly on the task execution. Using a hockey dribbling setting with experienced hockey players, \citet{ashford2010priming} present additional evidence for declining performance in pressure situations. However, \citeauthor{ashford2010priming} demonstrate that in a high-pressure priming condition, performances are equal to those in a low-pressure situation and better (thus faster) than in a high-pressure non-priming condition.

\citet{liao2002self} show decreasing free throw success for basketball novices in pressure situations. This result only applies to those subjects who are asked to pay close attention to the execution process during the practicing phase. Analyzing free throw performances of competitive basketball players instead of novices, \citet{wang2004self} find supporting results. Thus, participants suffer a significant decrease in free throw success when performing in a high-pressure situation induced by the introduction of an audience, videotaping and offering financial rewards for improved performance. 

\citet{mesagno2012choking} study the impact which the fear of negative evaluation has on choking under pressure. For short distance throwing of a basketball\footnote{Shots are taken from five different spots which all are placed at the distance of the free throw line.} under pressure conditions the authors only find decreasing performance (thus choking) for participants who were anxious about being evaluated negatively. For other subjects no significant difference in throwing success is found. 

Outside of experiments, field studies take advantage of the wealth of data on actual market participants who repeatedly perform almost identical tasks but under varying degrees of pressure. Pressure in these instances is determined by factors such as the importance of the competition considered, the current score in the competition, and the time left to play in a match.

Penalty kicks in soccer are considered to be a prototype pressure situation, as they critically affect the match outcome and the expectation to score a goal is very high. In line with the hypothesis of individuals tending to choke under pressure at skill tasks, \citet{dohmen2008professionals} finds success rates of penalty kicks in professional football to decline with increasing importance of success, i.e.\ as pressure increases. However, contradictory to these results, \citet{kocher2008performance} show success rates in penalty shootouts to increase with pressure in the German cup competition. \citet{kocher2012psychological} and \citet{apesteguia2010psychological} focus on the ``last-mover disadvantage'', i.e.\ whether teams that go first in a shootout have an advantage over the other team resulting from higher pressure from trailing. While \citet{apesteguia2010psychological} find that last-mover teams indeed suffer from this kind of pressure, \citet{kocher2012psychological} refute this finding and speculate the contradictory results to be a consequence of issues with the data considered. Potential reasons for varying success in penalty shootouts between players are covered by \citet{jordet2009why}, where it is found that players from high-status countries a) generally perform worse and b) engage more in escapist self-regulation strategies than players from low status-countries.

In golf, performance under pressure in analyzed for putting. \cite{clark2002evaluating,clark2002professional} analyzes the impact of the current scoreboard situation on performance and finds that interim results are irrelevant for performance. In particular players who are in the lead or close to the lead in the final round do not perform worse than those who are further behind. Furthermore, players' performances are constant across rounds. \citet{wells2012evidence} state that between-athlete comparisons may explain this finding, which is not in line with the widely accepted hypothesis of individuals choking under pressure. Considering also within-golfer comparisons, \citet{wells2012evidence} cannot replicate Clark's findings, and instead do find athletes to choke under pressure. Relating choking under pressure to golfers' age, \citet{fried2011impact} show an inverted U-shaped relationship on the professionals' tour with performance under pressure peaking at age 36. Finally, \citet{hickman2015impact} determine the success rate at the final putt of a golf tournament to decrease as the value associated with that shot increases. 

Basketball free throws constitute another scenario that is often considered to investigate performance under pressure. Considering data from the National Basketball Association (NBA), \citet{worthy2009choking} model free throw success rates as a function of the current score, focusing on closely contested games in the final minutes. While players are shown to perform much worse when their team either is trailing by 1 or 2 points, or in the lead with 1 point, more attempts are successful when the score is tied (which equals less pressure since a miss would end in an overtime and not a loss). \citet{cao2011performance} also report evidence for choking under pressure in professional basketball, with performance declining with additional pressure (specifically: less time remaining, closer games). However, they show performance to be unaffected by the crowd size, the tournament round, and whether or not it is a home game for the player considered. Examining the determinants of choking under pressure, \citet{toma2017missed} finds overall lower free-throw success rates for all groups considered, i.e.\ females and males, amateurs and professionals, in case of high-pressure situations.

While some contradictory results have been reported, overall there still seems to be fairly clear evidence that professional athletes do choke under pressure, at least in some scenarios. However, below we argue that many of the existing studies have major limitations, and hence question the generalization of the findings.

\subsection{Caveats of Existing Studies \& Data}
Despite the substantial effort that has gone into studying the impact of pressure of performance, here we question whether compelling evidence for choking under pressure has yet been found. Generally, empirical work often lacks a clear distinction between skill and effort tasks, which however is of crucial importance given that the expected impact of pressure is depending on the type of task at hand. More specifically, we see the following caveats in existing studies. 

First, data on penalty kicks in soccer involves the major disadvantage that the success probability depends on the performance of two interacting individuals. Thus, missing a penalty shot can be due to the kicker's or the goalkeeper's performance, respectively, or in fact both \citep{jordet2008avoidance,jordet2009why}. In order to precisely measure the impact of pressure, analyses need to focus on performance that is not affected by others \citep{baumeister1984paradoxical}. 
Second, experimental studies of choking under pressure typically ask subjects to perform tasks they are not trained in \citep{dohmen2008professionals,kocher2012psychological}, a circumstance that makes it difficult to describe a baseline performance level for these individuals in non-pressure situations, and subsequently uncover deviations in pressure situations. It also seems likely that untrained individuals will not exhibit the same reaction to pressure as highly skilled professionals. 
Third and closely related to the previous point, the tasks to be performed in a pressure situation are often unusual to the subject. This applies not only to experimental data, but also to data on penalty kicks --- as argued by \citet{feri2013there}, penalty kicks constitute only a small fraction of actions a soccer player need to perform. Consequently, the number of observations per subject --- if there are any --- is often very low, which is likely to lead to very noisy estimations \citep{gelman2018the}. As highlighted by \cite{beilock2007athletes}, suboptimal performance is not just a random fluctuation in performance, but the overall response to a high-pressure situation. Estimating skill levels in pressure situations requires the separation of signal and (potentially very large) noise, which is only possible provided that the data set considered is comprehensive. 
Fourth, with respect to penalty kicks as well as basketball free throws, team managers often rely on the same set of players when confronted with pressure situations, namely those who they have faith in to deal with the pressure \citep{dohmen2008professionals}. Accordingly, managers avoid making use of those players who choke under pressure, leading to endogeneity for team sport settings. 
Fifth and mainly concerning professional golf, in noisy settings, it is statistically challenging to separate the effect of pressure from nuisance effects such as wind or weather conditions, position of the ball, or quality of the course. 
Sixth, in sports like soccer and basketball, crucial situations occur late in the game when athletes have increased levels of fatigue. Corresponding investigations of performance under pressure thus focus on very specific settings, and it is not clear whether athletes that choke under pressure when exhausted would also do so under normal circumstances. 

Overall, while it is difficult to test for any actual bias that may have crept into existing studies, we believe that there is sufficient reason to doubt the generality of existing findings, and hence argue that further studies, with better data and less confounding factors, are required in order to arrive at conclusive results regarding the effect of pressure on performance.

\section{Pressure Situations in Darts}\label{chap:Pressure Situations in Darts} \label{s:pressuredarts}
Professional darts is a nearly optimal setting for investigating performance under pressure. Professional players repeatedly perform highly standardized actions, with no interference by an opponent or any teammates involved, and hardly any relevant external factors. Furthermore, throwing darts with high precision clearly is a skill task. Through years of practice the throwing motion becomes increasingly automatized. Performance then is no longer the result from a step-by-step process but rather from an automated sequence of movements. Considering the previous research in this area, we would therefore expect pressure to negatively affect performance. It is worth noting here that we focus on highly-trained athletes who deal with pressure on a regular basis, rather than amateurs in experimental settings \citep{hickman2015impact}. It is not clear \textit{per se} whether the effect of pressure is the same across such different types of individuals.

\begin{figure}[!htb]
\centering
\begin{tikzpicture}
\draw[-] (15,0) -- (0,0) node[below,pos=0.5]{};
\draw[<->] (12,-0.5) -- (0,-0.5) node[below,pos=0.5]{2.37m};
\draw[dashed, line width=0.6mm, blue] (12,0.4) -- (12,-0.4);
\node[text width=3cm] at (13, 0.75) {oche};
\draw[-,line width=0.9mm, red!80] (0.05, 9.75) -- (0.05, 7.75) node[right,pos=0.5,text=black]{dartboard};
\draw[-] (0,10) -- (0,0) node[left,pos=0.5]{};
\draw[<->] (-0.5,8.75) -- (-0.5,0) node[left,pos=0.5]{1.73m};
\end{tikzpicture}%
\caption{Player position relative to dartboard.} 
\label{fig:Dartmeasures}
\end{figure}
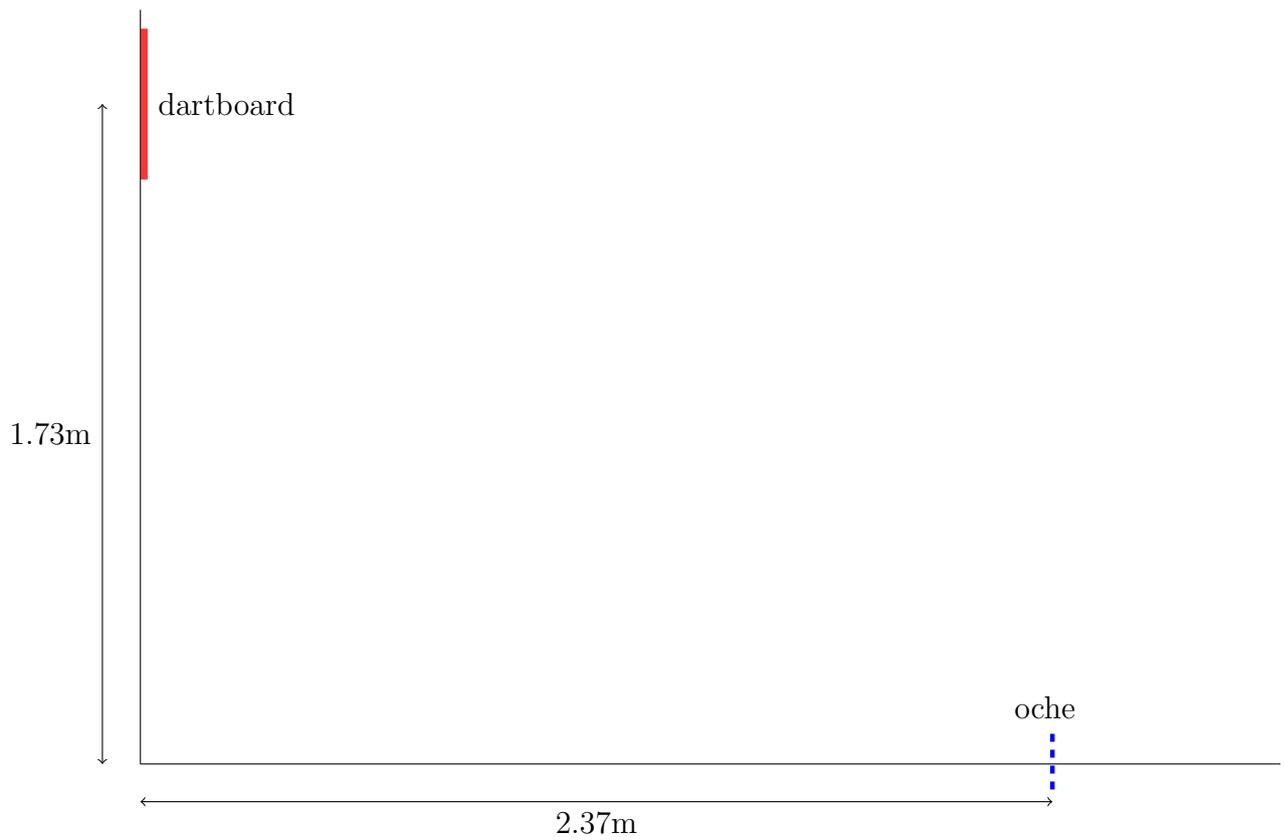

For readers who may be unfamiliar with the rules of darts, we here provide a short description. The dartboard consists of 20 different slices, which differ with respect to their value (ranging from 1 to 20), and the center of the board, which is composed of two fields, namely the single bull and the bullseye. Each slice is further divided into three different parts: two single, one double and one triple field. The bullseye is the double field of the single bull. Figure \ref{fig:Dartboard} shows the layout of a standard dartboard, highlighting the single five segment, the double and triple eight, respectively, and the single bull together with the bullseye. The inside width of the triple and double fields is 8mm, whereas the diameter of the bullseye is 12.7mm. A dart match is typically played by two players. (There are cases of team competitions in darts but these are not considered in our analysis.) Players are standing 2.37m away from the dartboard (at the ``oche''), the height of which is 1.73m (from the ground to the center of the bullseye). Figure \ref{fig:Dartmeasures} shows this setting. 

While there are many possible games in darts, professional darts commonly follow the \textit{501 up} format. In order to win a corresponding match, a player must be the first to win a pre-specified number of legs (typically between 7 and 15). Both players start each leg with 501 points, and the first player to reach exactly zero points wins the leg, with the restriction that the dart that ultimately reduces the points to zero must hit a double field. For instance, in case a player throws a dart at the single/double/triple field of segment 20, 20/40/60 points are deducted from the player's current score. The players take turns to throw three darts in quick succession. At the beginning of a leg, players consistently aim at high numbers --- usually triple 20 or triple 19 --- to quickly reduce their points. The maximum score per dart is 60 (triple 20) and hence 180 for a set of three darts.

\begin{figure}[!htb]
\centering
\begin{tikzpicture}
\node [anchor=west] (double field) at (-1,3) {triple field};
\node [anchor=west] (triple field) at (-1,1) {double field};
\node [anchor=east] (bullseye) at (13,2.8) {bullseye};
\node [anchor=east] (single bull) at (13.45, 6) {single bull};
\node [anchor=west] (singles) at (-1, 7) {single};
\begin{scope}[xshift=1.5cm]
    \node[anchor=south west,inner sep=0] (image) at (0,0) {\includegraphics[scale=0.55]{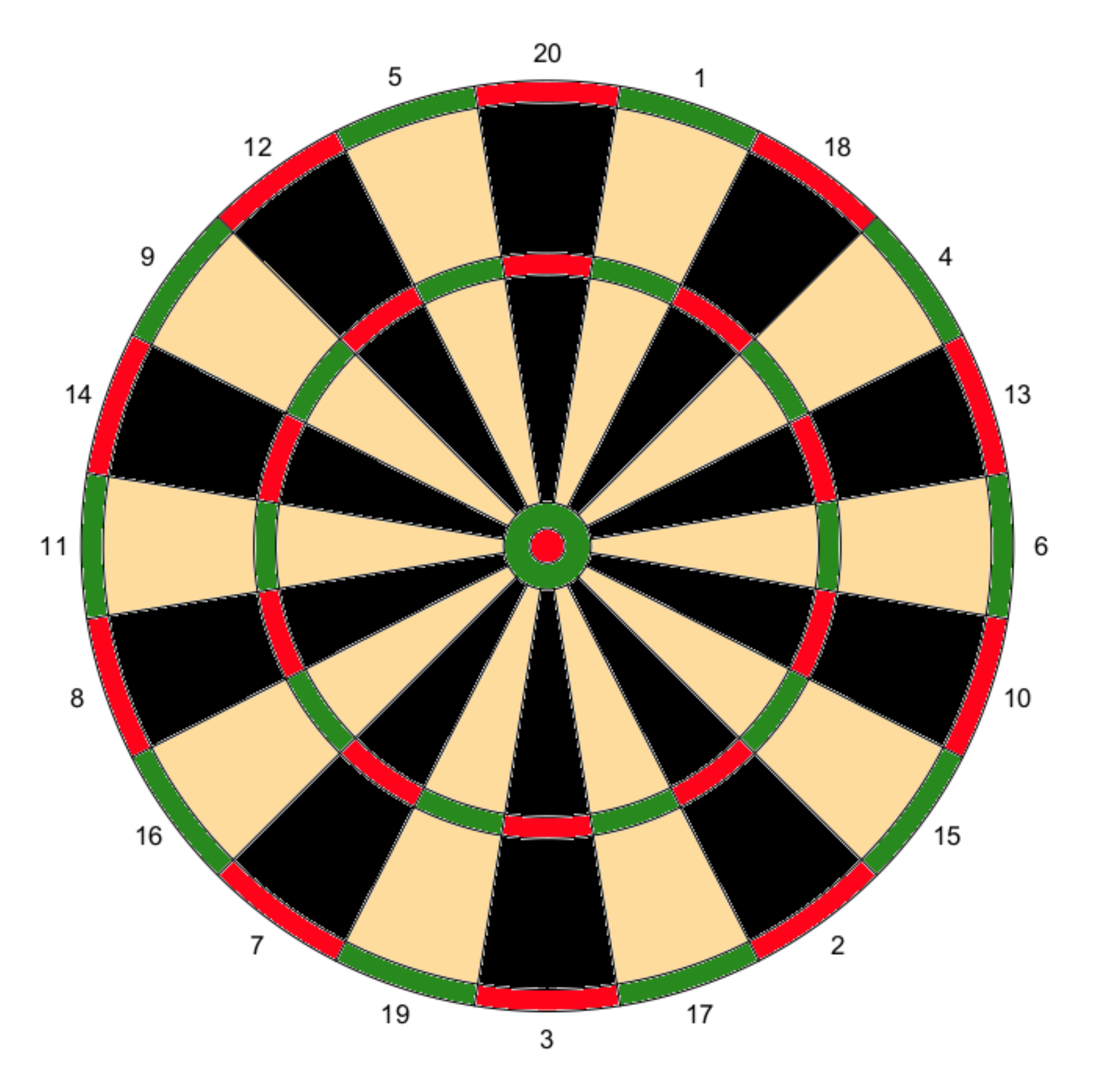}};
    \begin{scope}[x={(image.south east)},y={(image.north west)}]
        \draw [-stealth, line width=2pt, blue!80] (double field)  -- ++ (0.4 , 0.13);
        \draw [-stealth, line width=2pt, blue!80] (triple field) -- ++ (0.25, 0.25);
        \draw [-stealth, line width=2pt, blue!80] (bullseye) -- ++ (-0.59, 0.21);
        \draw [-stealth, line width=2pt, blue!80] (single bull) -- ++ (-0.59, -0.1);
        \draw [-stealth, line width=2pt, blue!80] (singles) -- ++ (0.6, 0.12);
        \draw [-stealth, line width=2pt, blue!80] (singles) -- ++ (0.63, -0.02);
    \end{scope}
\end{scope}
\end{tikzpicture}%
\caption{Dartboard layout.} 
\label{fig:Dartboard}
\end{figure}
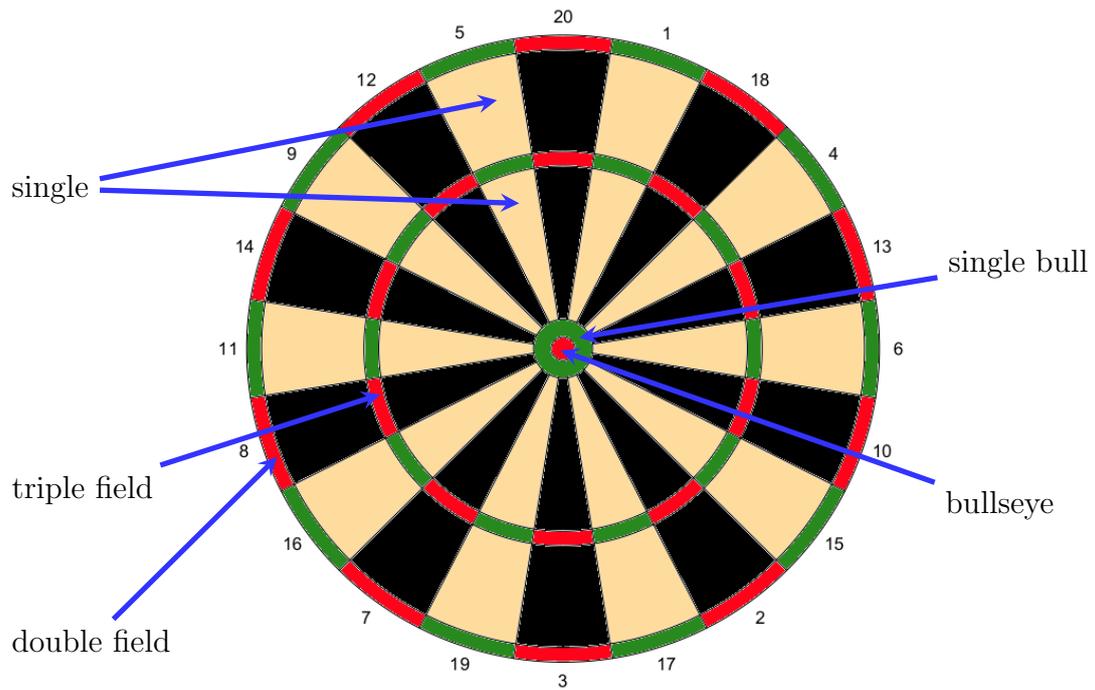

Once a player has the possibility to finish a leg (i.e.\ reach exactly zero points) with three darts (or less) during his turn, he is in the \textit{finish region}. If he takes the opportunity and finishes the leg, this is called a \textit{checkout}. As the last single dart has to hit a double field, the highest possible checkout is 170: two darts at triple 20 ($2 \times 60 = 120$) followed by a dart into the bullseye (50 points). The highest checkout not requiring a bullseye is 160 (two triple 20 followed by a double 20). For some scores below 170 there are multiple combinations for a checkout while there are none for others (e.g.\ 159 points as there is no three dart combination that leads to exactly zero points with the last dart hitting a double field\footnote{159 points could be reduced to exactly zero points with three darts if the last dart does not need to hit a double field, e.g.\ by triple 20 -- triple 20 -- triple 13. However, since all tournaments in our data are played as ``double out'', 159 points can not be reduced to zero within a players' turn.}). 

We classify how close a player is to checking out by differentiating between \textit{1-Dart-Finish}, \textit{2-Dart-Finish}, \textit{3-Dart-Finish} and \textit{no finish}. Within a \textit{1-Dart-Finish} the player is able to checkout with only one dart, e.g.\ in case he has 20 points left prior to his first throw he can checkout by hitting the double ten. Similarly, a \textit{2-Dart-Finish} describes a situation where two darts are needed to checkout, e.g.\ in case 58 points are left it is not possible to checkout with a single dart, but for example by a 20 followed by a double 19. Scores that need at least three darts for checkout are defined as a \textit{3-Dart-Finish}, for example 133 points left, which can be checked out via triple 20, triple 19 and double 8. \textit{No finish} describes situations in which it is not possible to checkout (e.g.\ because too many points are left or no three dart combination exists that leads to zero points).

The possibility to checkout represents the strongest pressure situation in darts, as in these situations a player can win a leg but also effectively lose a leg (if the opponent is likely to checkout in his next turn). For any given turn of a player, the level of pressure is a result from the combination of the player's own likelihood of finishing within the current turn as well as that of the opponent finishing within his next turn. Table \ref{tab:finishesByOpp2} summarizes the empirical proportions found in our data for winning a leg given the finish of a player (stated in the columns) and the finish of the opponent (stated in the rows). For a 1-Dart-Finish of a player, the situation where his opponent also has the possibility to checkout does not represent a situation of particularly high pressure, since the player has three darts to hit the targeted double. Furthermore, for this scenario, the empirical proportion for winning the leg is at least about 80\%. If a player has a 2-Dart-Finish, and in case his opponent has a 3-Dart-Finish (\textit{OppFin3}), he is still ahead of his opponent and will not feel much pressure, in relative terms, especially compared to the pressure situation where his opponent is in front, i.e.\ for the case where his opponent has a 1-Dart-Finish (\textit{OppFin1}). This classification of the pressure situations again occurs from the empirical proportions of winning a leg, as these decrease to 61.2\% for the mentioned scenario. The same argument applies for the 3-Dart-Finish of a player, where the pressure is highest for the situation where the opponent has a 1-Dart-Finish. In this situation, the empirical proportion of winning a leg is only 28.2\% and, hence, represents the situation with highest pressure. All possible combinations where the opponent has no finish (\textit{OppNoFin}) are situations where the player is not under pressure. Table \ref{tab:pressureclass} summarizes these different levels of pressure, where the levels of pressure can be directly deduced from the empirical proportions in Table \ref{tab:finishesByOpp2}. 
If players indeed choke under pressure one would assume a lower probability of checking out if the chances of the opponent to checkout with the next try increase.\footnote{For clarity imagine a situation where a player approaches the board with a \textit{3-Dart-Finish}. The pressure on him is lowest if his opponent is in the position of \textit{no finish} since he knows that his opponent cannot checkout in case he fails to finish himself and, hence, will have another turn after his current one. In comparison the pressure on the player is highest if his opponent has a \textit{1-Dart-Finish}. In case the player does not succeed, the chances of his opponent to win the leg with his next turn are very high (at around 75 percent in our sample).} For a given finish (\textit{1-Dart-Finish}, \textit{2-Dart-Finish} or \textit{3-Dart-Finish}) and in line with the choking under pressure literature we assume that a player's probability of checking out -- ceteris paribus -- increases in the number of darts needed by \textit{the opponent} to check out.

\begin{table}[ht]
\centering
 \caption{Empirical proportions of winning a leg under varying finish regions of player and opponent.} 
 \label{tab:finishesByOpp2}
  \begin{threeparttable}
\begin{tabular}{llll}
	\toprule
    & \textit{1-Dart-Finish} & \textit{2-Dart-Finish} & \textit{3-Dart-Finish} \\ 
	\midrule
    \textit{OppFin1} & 0.812 (2,368) & 0.612 (2,536) & 0.282 (1,902) \\ 
	\textit{OppFin2} & 0.854 (1,969) & 0.697 (2,700) & 0.415 (2,533) \\ 
	\textit{OppFin3} & 0.925 (936) & 0.845 (1,605) & 0.677 (2,250) \\ 
	\textit{OppNoFin} & 0.962 (496) & 0.938 (1,353) & 0.856 (2,544) \\ 
	\bottomrule 
\end{tabular} 
\begin{tablenotes} \footnotesize
\item \textit{Note}: number of observations in parentheses
\end{tablenotes}
\end{threeparttable}
\end{table}

\begin{table}[ht]
\centering
  \caption{Classification of pressure situations.} 
  \label{tab:pressureclass} 
\begin{tabular}{lccc}
  \hline
 & \textit{1-Dart-Finish} & \textit{2-Dart-Finish} & \textit{3-Dart-Finish} \\ 
  \hline
\textit{OppFin1} &  low pressure  & moderate pressure & high pressure \\ 
  \textit{OppFin2} & low pressure &  moderate pressure & moderate pressure \\ 
  \textit{OppFin3} & low pressure & low pressure & moderate pressure  \\ 
  \textit{OppNoFin} &  no pressure & no pressure &  no pressure \\ 
   \hline
\end{tabular}
\end{table}


\section{Empirical Analysis}\label{chap:Empirical Analysis}
The data -- extracted from \url{http://live.dartsdata.com/} -- covers all professional darts tournaments organized by the Professional Darts Corporation (PDC) between April 2017 and January 2018. Based on the raw data it was possible to reconstruct which player makes a throw, the score before each dart, how many legs have been played in the match, which player had the first throw in any leg considered and, of course, if the player making a throw checks out. In the data we analyze, each row, i.e.\ observation, corresponds to a player's turn to throw (at most) three darts. From those rows, i.e.\ from all sets of three darts played by a player, we consider only those instances where a player has the chance to check out within the given turn.\footnote{More specifically, the data indicate that for very few situations where a player has a 1-Dart-Finish and the opponent has no finish, the player does not always try to check out. For example, when having 10 points left, players sometimes go for the single two to have eight points left. Otherwise, if they throw at the double five and hit the single five, they have five points left which can not be reduced to zero with one dart as the last dart has to hit a double field. Hence, we excluded all throws where the player did not try to check out with his last dart, resulting in 11 removed observations.} To ensure reliable inference on player-specific effects, we further reduced the dataset to consider only those players who had at least 50 attempts to check out. The final data set comprises information on the checkout performances of $m=83$ different players, totaling to $n=23,192$ observations (checkout yes/no).

\subsection{Descriptive Statistics}
Our response variable \textit{Checkout} indicates whether a player managed to checkout (coded as ``Checkout$\,$=$\,$1'') or not (``Checkout$\,$=$\,$0''). As detailed in Section \ref{s:pressuredarts}, we measure the degree of pressure on a player by differentiating between his and his opponents' chances to finish a leg prior to his turn. The different ways to win a leg here refer to the minimum number of darts a player needs to checkout in his upcoming turn. We distinguish between a \textit{3-Dart-}, \textit{2-Dart-} or \textit{1-Dart-Finish} of the player. For the opponent, the covariates for the type of the finish are \textit{OppFin1}, \textit{OppFin2}, \textit{OppFin3} and \textit{OppNoFin}, indicating whether the opponent has a \textit{3-Dart-}, \textit{2-Dart-}, \textit{1-Dart-Finish} or \textit{no finish}, respectively. Finally, as our data contains trained athletes, we are able to further control for the experience of the athletes (\textit{Exper}), proxied by the number of years the players belong to a professional darts organization (BDO or PDC). Table \ref{tab:deskriptiv} summarizes all covariates considered.

\begin{table}[!htbp] \centering 
  \caption{Descriptive statistics for the covariates.} 
  \label{tab:deskriptiv} 
\begin{tabular}{lccccc}
\toprule
 & \textbf{Obs} & \textbf{Mean} & \textbf{Std.\ Dev.} & \textbf{Min} & \textbf{Max}\\ \midrule 
\textit{Checkout} & 23,192 & 0.390 & -- & 0 & 1 \\ 
\textit{OppNoFin} & 23,192 & 0.189 & -- & 0 & 1 \\ 
\textit{OppFin1} & 23,192 & 0.293 & -- & 0 & 1 \\ 
\textit{OppFin2} & 23,192 & 0.311 & -- & 0 & 1 \\ 
\textit{OppFin3} & 23,192 & 0.207 & -- & 0 & 1 \\ 
\textit{Exper} & 23,192 & 13.31 & 7.595 & 1 & 36 \\ 
\bottomrule
\end{tabular} 
\end{table}

\begin{table}[ht]
\centering
  \caption{Empirical proportions for a checkout under varying finish regions.} 
  \label{tab:finishes}
\begin{tabular}{lcc}
  \toprule
 & \textbf{Obs} & \textit{Checkout}$\,$=$\,$1 \\ 
  \midrule
 \textit{1-Dart-Finish} & 5,769 & 0.762 \\ 
  \textit{2-Dart-Finish} & 8,194 & 0.471 \\ 
  \textit{3-Dart-Finish} & 9,229 & 0.085 \\  
  \midrule
  \textbf{total} & 23,192 & 0.390 \\  
   \bottomrule
\end{tabular}
\end{table}

As can be seen in Table \ref{tab:finishes}, about 39\% of all finishes are successful checkouts. However, the probability to successfully complete a checkout varies considerably with the number of darts required to do so: the more darts are needed, the less likely is a checkout. At this point still neglecting the opponents' score, players check out when having a 3-Dart-Finish in only about 9\%, when having a 2-Dart-Finish in about 47\% and when having a 1-Dart-Finish in about 76\% of all cases.

\begin{table}[ht]
\centering
 \caption{Empirical checkout proportions under varying finish regions of player and opponent.} 
 \label{tab:finishesByOpp}
  \begin{threeparttable}
\begin{tabular}{llll}
	\toprule
    & \textit{1-Dart-Finish} & \textit{2-Dart-Finish} & \textit{3-Dart-Finish} \\ 
	\midrule
    \textit{OppFin1} & 0.757 (2,368) & 0.496 (2,536) & 0.114 (1,902) \\ 
	\textit{OppFin2} & 0.763 (1,969) & 0.478 (2,700) & 0.093 (2,533) \\ 
	\textit{OppFin3} & 0.770 (936) & 0.447 (1,605) & 0.080 (2,250) \\ 
	\textit{OppNoFin} & 0.770 (496) & 0.435 (1,353) & 0.061 (2,544) \\ 
	\bottomrule 
\end{tabular} 
\begin{tablenotes} \footnotesize
\item \textit{Note}: number of observations in parentheses
\end{tablenotes}
\end{threeparttable}
\end{table}

Table \ref{tab:finishesByOpp} additionally includes information on opponents' possibilities to finish a leg. It shows all combinations of players' and the opponents' finish regions together with the corresponding empirical proportions of successful checkouts. The corresponding number of observations for each situation is shown in parentheses. For a \textit{1-Dart-Finish}, the checkout proportions do not vary considerably across the opponents' finish region. In contrast, for both the \textit{2-Dart-Finish} and also the \textit{3-Dart-Finish}, there are notable differences in the checkout proportions across the different finish regions of the opponent. For \textit{2-Dart-Finishes}, the checkout proportion increases as the opponent's chance to successfully complete a checkout is increased, i.e.\ for situations associated with higher pressure. The same holds for \textit{3-Dart-Finishes}, since the checkout proportion almost doubles from situations where the opponent cannot checkout and hence does not apply much pressure (0.061) to the situation where the opponent has a \textit{1-Dart-Finish}, i.e.\ situations of high pressure (0.114). 

\subsection{Modelling Checkout Performance}
The structure of the data considered is longitudinal, as we model the binary response variable \textit{Checkout}$_{ij}$, indicating whether or not the $i$--th player ($i=1,\ldots,m$) checked out (\textit{Checkout}$_{ij}=1$) on the $j$--th attempt ($j=1,\ldots,n_i$). To cover player-specific effects, and also to account for the fact that each individual player's observations are likely to be correlated, we apply generalised linear mixed models where the linear predictor $\eta_{ij}$ contains a vector of fixed effects $\boldsymbol{\beta}$ as well as a vector of zero-mean random effects $\boldsymbol{\gamma}_i$:
\begin{equation*}
\eta_{ij} = \boldsymbol{x}_{ij}' \boldsymbol{\beta} + \boldsymbol{u}_{ij}'\boldsymbol{\gamma}_i, \qquad i=1, \dots, m, \quad j=1,\dots,n_i,
\end{equation*}
with $\boldsymbol{x}_{ij} = (1, \textit{OppFin1}_{ij}, \ldots)'$, and $\boldsymbol{u}_{ij}'$ the subvector of $\boldsymbol{x}_{ij}'$ with those covariates for which we assume individual-specific effects. The logit function links the binary response variable, \textit{Checkout}$_{ij}$, to the linear predictor:
\begin{equation*}
\text{logit} \bigl( \Pr (\textit{Checkout}_{ij}=1 | \boldsymbol{\gamma}_{i}) \bigr) 
= \eta_{ij} = \boldsymbol{x}_{ij}' \boldsymbol{\beta} + \boldsymbol{u}_{ij}'\boldsymbol{\gamma}_i.
\end{equation*}
The linear predictor for \textit{Model 1} includes all covariates considered as well as a random intercept for each player to account for player-specific effects:
\begin{equation*}
\begin{split}
\eta_{ij} &  = \beta_0 + 
 \beta_1 \textit{OppFin1}_{ij} + \beta_{2} \textit{OppFin2}_{ij} + \beta_{3} \textit{OppFin3}_{ij} +  \beta_4 \textit{Exper}_{i} + \gamma_{0i}.
\end{split}
\label{Modell1}
\end{equation*}
The random intercept $\gamma_{0i}$ displays the player-specific deviation from the average intercept $\beta_0$ --- further individual-specific effects will be considered below. For the covariates \textit{OppFin1}/\textit{OppFin2}/\textit{OppFin3}, the reference category is \textit{OppNoFin}. 
Since the descriptive analysis showed that the magnitude of the effect of the finish of the opponent is likely to vary across the player's finish considered, we estimate \textit{Model 1} three times separately using only 1-Dart, 2-Dart and 3-Dart-Finish data. These models are fitted by maximum likelihood estimation using the package \texttt{lme4} (\citealt{lme4}) in R \citep{RCoreteam}. Table \ref{tab:modsep1} displays the results for the corresponding fixed effects.

\begin{table}[!htbp] \centering 
  \caption{Estimation results for the fixed effects of \textit{Model 1}.} 
  \label{tab:modsep1} 
  \scalebox{0.75}{
\begin{tabular}{@{\extracolsep{5pt}}lccc} 
\\[-1.8ex]\hline 
\hline \\[-1.8ex] 
 & \multicolumn{3}{c}{\textit{Response variable:}} \\ 
\cline{2-4} 
\\[-1.8ex] & \multicolumn{3}{c}{\textit{Checkout}} \\ 
\\[-1.8ex] & 1-Dart-Finish & 2-Dart-Finish & 3-Dart-Finish\\ 
\hline \\[-1.8ex] 
 \textit{OppFin1} & $-$0.067 & 0.267$^{***}$ & 0.684$^{***}$ \\ 
  & (0.117) & (0.068) & (0.110) \\ 
  & & & \\ 
 \textit{OppFin2} & $-$0.038 & 0.184$^{***}$ & 0.451$^{***}$ \\ 
  & (0.119) & (0.067) & (0.108) \\ 
  & & & \\ 
 \textit{OppFin3} & 0.002 & 0.057 & 0.285$^{**}$ \\ 
  & (0.132) & (0.075) & (0.114) \\ 
  & & & \\ 
   \textit{OppNoFin} &  \multicolumn{3}{c}{\textit{reference category}}  \\ 
  &  &  &  \\   \hdashline
  & & & \\ 
 \textit{Exper} & 0.0004 & 0.008$^{*}$ & 0.010 \\ 
  & (0.005) & (0.004) & (0.006) \\ 
  & & & \\ 
 \textit{Constant} & 1.186$^{***}$ & $-$0.424$^{***}$ & $-$2.916$^{***}$ \\ 
  & (0.125) & (0.080) & (0.124) \\ 
  & & & \\ 
\hline \\[-1.8ex] 
Observations & 5,769 & 8,194 & 9,229 \\ 
AIC & 6,334 & 11,294 & 5,341 \\ 
\hline 
\hline \\[-1.8ex] 
\textit{Note:}  & \multicolumn{3}{r}{$^{*}$p$<$0.1; $^{**}$p$<$0.05; $^{***}$p$<$0.01} \\ 
\end{tabular}} 
\end{table} 

The estimated coefficients associated with \textit{OppFin1}, \textit{OppFin2} and \textit{OppFin3} are of main interest here as they display the impact of the opponent's chance of checking out during his next attempt. For the 1-Dart-Finish of a player, the pressure situation has no statistically significant effect on the checkout. However, for the 2-Dart-Finish as well as the 3-Dart-Finish, the various effects of the dummy variables indicating whether the opponent has a finish are all estimated to be positive, and most are statistically significant. For example, when a player has a 2-Dart-Finish, then a 1-Dart-Finish of the opponent --- all other covariates held constant --- increases the odds of checking out by $\text{exp}(0.267) = 1.31$ relative to a situation where the opponent has no finish. 
Perhaps somewhat surprisingly, evaluating the effects of \textit{OppFin1}, \textit{OppFin2} and \textit{OppFin3} across the three models, the more pressure a player is exposed to, i.e.\ the less number of darts the opponent needs for a checkout, the higher is the increase in the corresponding odds for a checkout. 
Regarding the 1-Dart-Finish, players generally do not feel much pressure as they have three darts to hit the required double field. For this situation, we do not observe any significant effect of the \textit{OppFin1}, \textit{OppFin2} and \textit{OppFin3} covariates, respectively. For a 2-Dart-Finish, players are under little pressure if their opponent has a 3-Dart-Finish, since it is likely that they will have another turn to check out, as the opponent has to hit the targeted three fields perfectly in order to win the leg in his next turn (and note the empirical proportion for checking out when having a 3-Dart-Finish is about 0.085, see Table \ref{tab:finishes}). However, if the opponent has a 1-Dart-Finish or a 2-Dart-Finish, then players are exposed to more pressure as they know that for these situations it is more likely that their opponent will check out during his next turn. For these pressure situations, the odds for a checkout increase compared to a situation where the opponent does not have the possibility to check out. 
For a 3-Dart-Finish of any player considered, the same arguments apply, with the main difference being that the pressure is higher overall, since all 3 darts (rather than just 2) have to find their target. 
Notably, the estimated effects are in contrast to the widely accepted theory regarding choking under pressure, as discussed above.

\begin{figure}[!htb]
\centering
\includegraphics[scale=0.8]{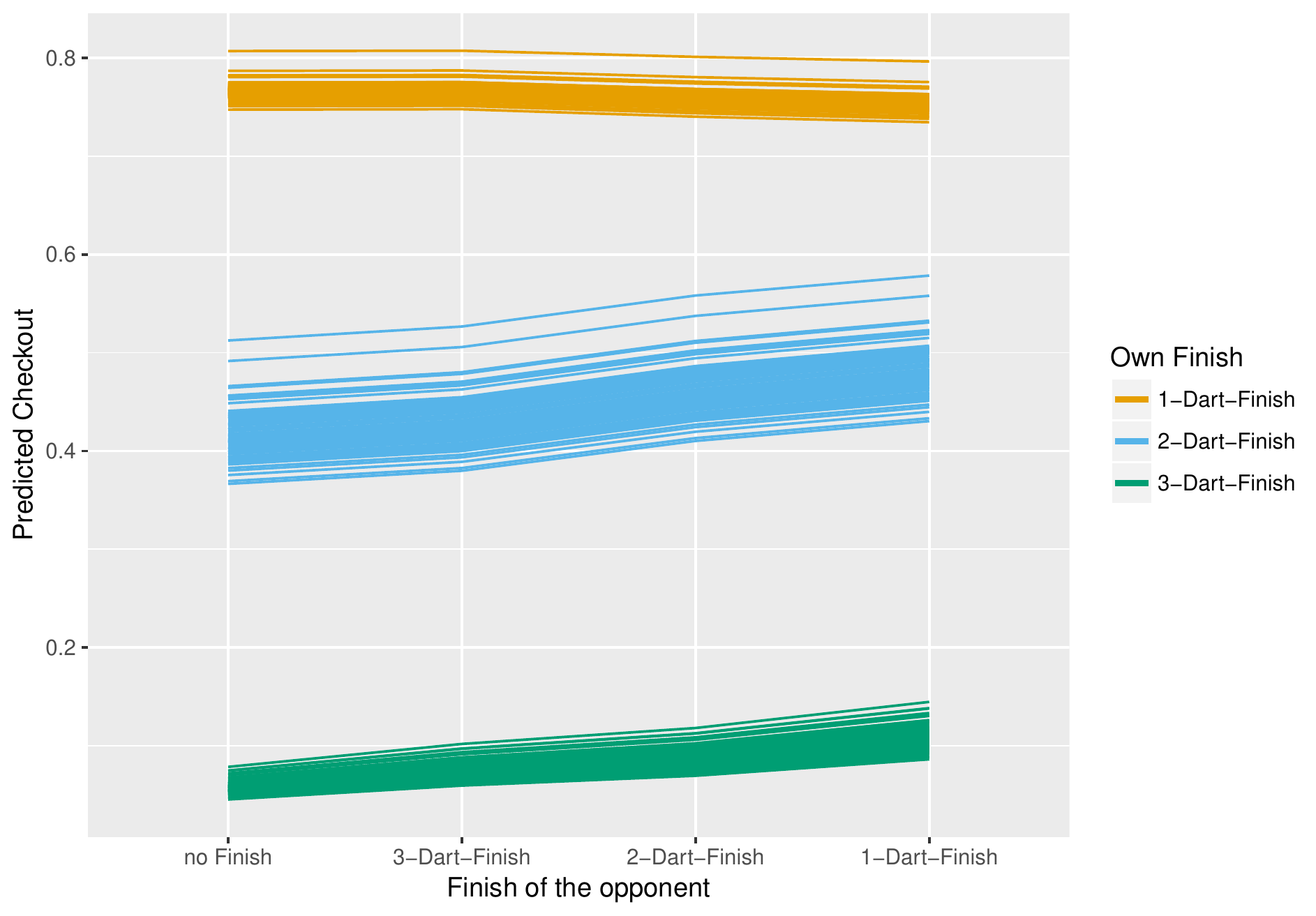}
\caption{Predicted \textit{Checkout} values according to \textit{Model 1}. More specifically, for each of the 83 players we predict the \textit{Checkout} for all combinations of the own finish (yellow, blue and green lines) and the finish of the opponent.} 
\label{fig:PredCheckout}
\end{figure}

The player-specific random intercepts $\hat{\gamma}_{0i}$, i.e.\ the player-specific deviations from the intercept $\hat{\beta}_0$, vary (on the logistic scale) between $-0.104$ and $0.239$ for the 1-Dart-Finish, between $-0.208$ and $0.381$ for the 2-Dart-Finish and between $-0.200$ and $0.383$ for the 3-Dart-Finish, respectively. To provide a better understanding and intuition of the player-specific effects, and of all three estimations of \textit{Model 1} in general, Figure \ref{fig:PredCheckout} shows the predicted values for \textit{Checkout} for each combination of the players' finish and the one of the opponent.
Thus, for each player there is one line for each of his own possible finishes (indicated by the three different colors), showing the predicted probability of finishing a leg as a function of the number of darts needed by the opponent in his next attempt. Thus, Figure \ref{fig:PredCheckout} summarizes several aspects of the fitted models as reported in Table \ref{tab:modsep1}: first, as indicated by the player-specific random intercepts, there are huge differences between the players' checkout performances; second, for the 1-Dart-Finish of a player, the checkout probability is not affected by the finish situation of the opponent; and third, for the 2-Dart-Finish as well as the 3-Dart-Finish of a player, the probability for checking out increases as the likelihood of a successful finish of the opponent increase, i.e.\ the more a player is under pressure.

Figure \ref{fig:PredCheckout} also highlights an implicit assumption of \textit{Model 1} namely that the effect of the dummies \textit{OppFin1}, \textit{OppFin2} and \textit{OppFin3} is the same for each player. To investigate whether the 83 players in the data to indeed respond in the same way to pressure situations, we extended \textit{Model 1} to include additional zero-mean random effects, $\gamma_{1i}, \gamma_{2i}$ and $\gamma_{3i}$, which represent the player-specific deviations from the fixed effects $\beta_1, \beta_2$ and $\beta_3$, leading to \textit{Model 2}: 
\begin{equation*}
\begin{split}
\eta_{ij} &  = \beta_0 + 
 \beta_1  \textit{OppFin1}_{ij} + \beta_{2} \textit{OppFin2}_{ij} + \beta_{3} \textit{OppFin3}_{ij} + \beta_4 \textit{Exper}_{i}\\ 
 &  +\gamma_{0i} + \gamma_{1i} \textit{OppFin1}_{ij} + \gamma_{2i} \textit{OppFin2}_{ij} + \gamma_{3i} \textit{OppFin3}_{ij}.
\end{split}
\label{Modell2}
\end{equation*}
As for \textit{Model 1}, we estimated \textit{Model 2} for each subsample of possible finishes separately. Table \ref{tab:modsep2} display the corresponding results,
indicating that the fixed effects change only slightly compared to \textit{Model 1}, both in terms of the direction of the effect and also its magnitude. The estimated random effects $\hat{\gamma}_{1i}, \hat{\gamma}_{2i}$ and $\hat{\gamma}_{3i}$ imply that for 2-Dart and 3-Dart-Finishes of a player, only two players do not improve if their opponent has a 2-Dart-Finish (compared to a situation where the opponent has no finish), and all players improve if the opponent has a 1-Dart-Finish, respectively. Table \ref{tab:pressuresum} summarizes these results and displays how many out of the 83 players in our sample improve their odds for a checkout when facing pressure situations. Thus, the empty cells in Table \ref{tab:pressuresum} display the non-pressure situations which were pointed out above. While \textit{Model 2} provides some insights regarding player-specific performances under pressure, \textit{Model 2} does not yield an improvement in the AIC compared to \textit{Model 1} for any of the subsamples modeled.

\begin{table}[!htbp] \centering 
  \caption{Estimation results for the fixed effects of \textit{Model 2}.} 
  \label{tab:modsep2} 
  \scalebox{0.75}{
\begin{tabular}{@{\extracolsep{5pt}}lccc} 
\\[-1.8ex]\hline 
\hline \\[-1.8ex] 
 & \multicolumn{3}{c}{\textit{Response variable:}} \\ 
\cline{2-4} 
\\[-1.8ex] & \multicolumn{3}{c}{\textit{Checkout}} \\ 
\\[-1.8ex] & 1-Dart-Finish & 2-Dart-Finish & 3-Dart-Finish\\ 
\hline \\[-1.8ex] 
 \textit{OppFin1} & $-$0.118 & 0.279$^{***}$ & 0.648$^{***}$ \\ 
  & (0.150) & (0.077) & (0.123) \\ 
  & & & \\ 
 \textit{OppFin2} & $-$0.089 & 0.226$^{***}$ & 0.454$^{***}$ \\ 
  & (0.149) & (0.079) & (0.115) \\ 
  & & & \\ 
 \textit{OppFin3} & $-$0.050 & 0.086 & 0.254$^{**}$ \\ 
  & (0.159) & (0.088) & (0.125) \\ 
  & & & \\ 
     \textit{OppNoFin} &  \multicolumn{3}{c}{\textit{reference category}}  \\ 
  &  &  &  \\   \hdashline
  & & & \\ 
 \textit{Exper} & 0.001 & 0.008$^{*}$ & 0.011$^{*}$ \\ 
  & (0.005) & (0.004) & (0.006) \\ 
  & & & \\ 
 \textit{Constant} & 1.227$^{***}$ & $-$0.450$^{***}$ & $-$2.913$^{***}$ \\ 
  & (0.153) & (0.087) & (0.127) \\ 
  & & & \\ 
\hline \\[-1.8ex] 
Observations & 5,769 & 8,194 & 9,229 \\ 
AIC & 6,349 & 11,304 & 5,358 \\ 
\hline 
\hline \\[-1.8ex] 
\textit{Note:}  & \multicolumn{3}{r}{$^{*}$p$<$0.1; $^{**}$p$<$0.05; $^{***}$p$<$0.01} \\ 
\end{tabular}}
\end{table} 

\begin{table}[ht]
\centering
  \caption{Number of players (out of $n=83$) that improve under pressure according to \textit{Model 2}.}
  \label{tab:pressuresum} 
\begin{tabular}{lccc}
  \toprule
 & \textit{1-Dart-Finish} & \textit{2-Dart-Finish} & \textit{3-Dart-Finish} \\ 
  \midrule
\textit{OppFin1} & -- & 83 & 83 \\ 
  \textit{OppFin2} & -- & 81 & 83 \\ 
  \textit{OppFin3} & -- & -- & 83 \\ 
   \bottomrule
\end{tabular}
\end{table}

\section{Throwing Darts --- Skill or Effort Task?}\label{chap:Throwing Darts - Skill or Effort Task?}
As pointed out above, throwing darts reflects a skill and not an effort task. It takes not much effort to throw the light-weighted dart towards the dart board. However, a high level of concentration and calm are needed in order to be successful. Because the distinction between effort and skill task is critical to the expected impact of pressure on performance, this section empirically tackles if incentives that typically impact effort also determine performance in this skill-based setting. Hence we enrich the model presented in the previous section by factors that would impact performance if the task at hand was an effort task.

If throwing darts was an effort task, the following results derived from contest theory would hold. First, costly effort and hence performance would improve as financial incentives increase \citep[see e.g.][]{becker1992incentive,ehrenberg1990tournaments,gilsdorf2008testing,gilsdorf2008tournament,lynch2005effort}. \textit{Prize} displays the additional money to be won in a game, i.e.\ the difference between the winners' and the losers' price money. If throwing darts was an effort task, then the checkout probability should increase in \textit{Prize}. Second, heterogeneity in ability between contestants would reduce performance for effort tasks as the outcome of the contest is more certain in advance and both contestants save effort costs \citep[see e.g.][]{bach2009incentive,backes2013tournament,brown2011quitters,sunde2009heterogeneity}. To account for the homogeneity of players chances to win the match, the competitive imbalance (\textit{Cb}) indicates the absolute difference in the winning probabilities. Based on betting odds from \url{http://www.oddsportal.com/}, and after correcting for the bookmakers' margin, they can take values between 0 and 1. High values of \textit{Cb} imply that the match is lopsided, whereas the value 0 means that both players have equal winning probabilities. One would expect a higher \textit{Cb} to be associated with lower incentives to perform well, hence a negative impact on the probability to check out. Third and closely related, intermediate scores suggesting a clear lead would have similar impact on effort and hence performance as heterogeneity between contestants abilities 
\citep[see e.g.][]{azmat2010importance,casas2009relative,gurtler2010feedback,schneemann2017intermediate}. Here, intermediate scores indicating an asymmetric contest reduce incentives to invest effort and hence reduces performance. The variable \textit{Diff} covers the difference in legs won at the time a dart is thrown. The larger \textit{Diff}, the more lopsided the course of play. Accordingly, if the task at hand was an effort task one would expect a negative impact of \textit{Diff} on the probability of a successful checkout. Last, if throwing darts was an effort task, one would expect a decline in performance as the game progresses and athletes fatigue. \textit{ThrowNumber} indicates how many throws a player had made prior to the current throw. 

The descriptive statistics for the effort-related control variables are displayed in Table \ref{tab:deskriptiv2} in the Appendix, while the estimation results covering those additional control variables are presented in Table \ref{tab:othercovariates}. The results for the considered covariates show mostly insignificant effects of the effort-related control variables on the checkout probability. Only for the difference in the score of the match we find mixed evidence but no clear trend.

\section{Discussion}\label{chap:discussion}
We provide clear evidence that professional darts players do not choke but instead excel when facing (high) pressure situations. Player-specific effects for performance under pressure in our models show that almost all professional players in our sample improve their overall performance in pressure situations. Our results contradict the current state of research, where it is widely accepted that overall performance in skill tasks decreases with increasing pressure due to choking. 

The stark difference between our findings and previous studies may partly be due to the fact that in our study we consider very highly skilled individuals who have to deal with the considered type of pressure situations on a regular basis. Professional darts players are at the very top of their profession and cannot fluke out of pressure situations, which is possible in team settings where tasks can be assigned to different team members. In fact, darts players face pressure situations on a regular basis and hence gain experience in dealing with these. While throwing darts is the one skill required in the setting considered, in other professions the set of tasks is much more diverse. With respect to previous studies, e.g.\ free throws in basketball or penalty kicks in soccer display only a very small fraction of the skills required of a player. Here, professionals can compensate weaknesses (like choking under pressure) with other strengths. 

While nearly all subjects in our sample respond positively to pressure, this does not necessarily imply similar results for (less experienced) subjects outside of the sample \citep{kamenica2012behavioral}. Instead, our sample may to some extent be the result of selection effects of subjects who can withstand pressure, such that only those individuals who perform well in pressure situations succeeded in the profession at hand and made it to the top (and hence into our sample). 

The importance of coping with pressure situations has been investigated by \citet{jones2002thing} in a qualitative study by interviewing ten international top athletes. In his study, several attributes are stated as important factors for being ``mental tough'', such as to be in control under pressure. In a further study, \citet{jones2007framework} again interview several former Olympic or world championship winning athletes as well as sport psychologists and coaches, finding that mentally tough athletes can not only cope with pressure situations, but even use it to raise their performance. \citet{jones2009theory} deliver an explanation for that, stating that individuals are either entering a ``competition state'' or a ``threat state'' when forced to pressure situations, where the former helps their performance and the latter does not. Thus, to not choke under pressure is not a conscious decision but rather a state of mind which is reached subconsciously. Our results suggest that the ability to enter a state of mind which is associated with a high level of focus when facing pressure situations (the ``competition state'') may be necessary to become one of the best at a profession.



\section{Appendix}

\begin{table}[!htbp] \centering 
  \caption{Descriptive statistics for the further considered covariates for the models shown in Table \ref{tab:othercovariates}.} 
  \label{tab:deskriptiv2} 
\begin{tabular}{@{\extracolsep{5pt}}lccccc} 
\\[-1.8ex]\hline 
\hline \\[-1.8ex] 
 & \multicolumn{1}{c}{n} & \multicolumn{1}{c}{mean} & \multicolumn{1}{c}{st.\ dev.} & \multicolumn{1}{c}{min.} & \multicolumn{1}{c}{max.} \\ 
\hline \\[-1.8ex] 
\textit{ThrowNumber} & 23,192 & 37.35 & 28.52 & 3 & 253 \\ 
\textit{Prize} & 21,601 & 7,739 & 17,750 & 1,000 & 230,000 \\ 
\textit{Cb} & 23,192 & 0.377 & 0.231 & 0.000 & 0.899 \\ 
\textit{Diff} & 19,088 & 1.497 & 1.452 & 0 & 11 \\ 
\hline \\[-1.8ex] 
\end{tabular} 
\end{table}

\begin{table}[!htbp] \centering 
\caption{Estimation results for the fixed effects of the models containing further covariates. The number of observations differ, since some darts tournaments such as the Premier League start with a group stage where no prize money is awarded. In addition, the World Championship is played in sets, which makes the intermediate score incomparable with tournaments played in legs.}
 \label{tab:othercovariates} 
\begin{tabular}{@{\extracolsep{5pt}}lccc} 
\\[-1.8ex]\hline 
\hline \\[-1.8ex] 
 & \multicolumn{3}{c}{\textit{Response variable:}} \\ 
\cline{2-4} 
\\[-1.8ex] & \multicolumn{3}{c}{\textit{Checkout}} \\ 
\\[-1.8ex] & 1-Dart-Finish & 2-Dart-Finish & 3-Dart-Finish \\ 
\hline \\[-1.8ex] 
 \textit{OppFin1} & $-$0.121 & 0.294$^{***}$ & 0.681$^{***}$ \\ 
  & (0.136) & (0.078) & (0.129) \\ 
  & & & \\ 
 \textit{OppFin2} & $-$0.059 & 0.183$^{**}$ & 0.444$^{***}$ \\ 
  & (0.139) & (0.077) & (0.126) \\ 
  & & & \\ 
 \textit{OppFin3} & $-$0.036 & 0.132 & 0.333$^{**}$ \\ 
  & (0.152) & (0.086) & (0.131) \\ 
  & & & \\ 
     \textit{OppNoFin} &  \multicolumn{3}{c}{\textit{reference category}}  \\ 
  &  &  &  \\   \hdashline
  & & & \\ 
 \textit{Exper} & 0.001 & 0.056 & 0.077 \\ 
  & (0.041) & (0.035) & (0.054) \\ 
  & & & \\ 
 \textit{Cb} & $-$0.005 & 0.033 & 0.089$^{*}$ \\ 
  & (0.039) & (0.029) & (0.048) \\ 
  & & & \\ 
 \textit{ThrowNumber} & $-$0.069 & 0.045 & $-$0.009 \\ 
  & (0.048) & (0.035) & (0.058) \\ 
  & & & \\ 
 \textit{Prize} & 0.023 & 0.041 & $-$0.045 \\ 
  & (0.066) & (0.048) & (0.086) \\ 
  & & & \\ 
 \textit{Diff} & 0.021 & $-$0.059$^{**}$ & $-$0.013 \\ 
  & (0.038) & (0.028) & (0.046) \\ 
  & & & \\ 
 Constant & 1.232$^{***}$ & $-$0.318$^{***}$ & $-$2.803$^{***}$ \\ 
  & (0.127) & (0.070) & (0.105) \\ 
  & & & \\ 
\hline \\[-1.8ex] 
Observations & 4,409 & 6,172 & 6,916 \\ 
AIC & 4,845 & 8,509 & 3,961 \\ 
\hline 
\hline \\[-1.8ex] 
\textit{Note:}  & \multicolumn{3}{r}{$^{*}$p$<$0.1; $^{**}$p$<$0.05; $^{***}$p$<$0.01} \\ 
\end{tabular} 
\end{table}

\clearpage

\bibliographystyle{apalike}
\bibliography{bibstrategy}

\clearpage

\end{document}